# Manipulation of Charge Delocalization in a Bulk Heterojunction Material Using a Mid-Infrared Push Pulse

Angela Montanaro[1,2,·], Kyu Hyung Park[3,·], Francesca Fassioli[2,3,4], Francesca Giusti[1], Daniele Fausti[*,1,2], and Gregory D. Scholes[*,3]

[1] Department of Physics, University of Trieste, Via A. Valerio 2, 34127 Trieste, Italy; Elettra-Sincrotrone Trieste S.C.p.A. Strada Statale 14 - km 163.5 in AREA Science Park 34149 Basovizza, Trieste, Italy

[2] Department of Physics, University of Erlangen-Nürnberg, 91058 Erlangen, Germany

[3] Department of Chemistry, Princeton University, Princeton, New Jersey 08544, United States

[4] SISSA − Scuola Internazionale Superiore di Studi Avanzati, Trieste 34136, Italy

· These authors contributed equally.

* Corresponding authors: gscholes@princeton.edu; daniele.fausti@elettra.eu

## ABSTRACT

In organic bulk heterojunction materials, charge delocalization has been proposed to play a vital role in the generation of free carriers by effectively reducing the Coulomb attraction via an interfacial charge transfer exciton (CTX). Pump-push-probe (PPP) experiments produced evidence that the excess energy given by a push pulse enhances delocalization, thereby increasing photocurrent. However, previous studies have employed near-infrared push pulses in the range ~0.4-0.6 eV which is larger than the binding energy of a typical CTX. This raises the doubt that the push pulse may directly promote dissociation without involving delocalized states. Here, we perform PPP experiments with mid-infrared push pulses at energies that are well below the binding energy of a CTX state (0.12-0.25 eV). We identify three types of CTX: delocalized, localized, and trapped. The excitation resides over multiple polymer chains in delocalized CTXs, while is restricted to a single chain (albeit maintaining a degree of intrachain delocalization) in localized CTXs. Trapped CTXs are instead completely localized. The pump pulse generates a "hot" delocalized CTX, which promptly relaxes to a localized CTX, and eventually to trapped states. We find that photo-exciting localized CTXs with push pulses resonant to the mid-infrared charge transfer absorption can promote delocalization and, in turn, contribute to the formation of long-lived charge separated states. On the other hand, we found that trapped CTX are non-responsive to the push pulses. We hypothesize that delocalized states identified in prior studies are only accessible in systems where there is significant interchain electronic coupling or regioregularity that supports either interchain or intrachain polaron delocalization. This, in turn, emphasizes the importance of engineering the micromorphology and energetics of the donor-acceptor interface to exploit a full potential of a material for photovoltaic applications.

## INTRODUCTION

The near unity internal quantum efficiency (IQE) of charge generation in organic solar cell materials [1-4] is often ascribed to the peculiar nature of interface charge transfer states. A key role is played by states known as charge transfer excitons (CTXs) which are described as a superposition of a neutral exciton and bound singly charged polaron pairs. The mystery of how CTXs in a low dielectric environment overcome their large binding energies has bred a wealthy of discussion on its working mechanism. While the role of entropy and energetic disorder has been invoked to play a decisive role in the efficient dissociation of the CTXs into free charges [5-8], one of the most plausible hypotheses to explain the efficient dissociation of the CTXs into free charges is a "hot" state model, which states that higher vibrational or electronic states of CTX, populated by an initially generated singlet excitons ($^1$Ex) with excess energy, can easily cross the Coulomb barrier of charge separation [9].

While a large body of literature has been devoted towards supporting this view, recent evidence is often against it [10-14]. Such objections are based on the measurement of device efficiencies with systematic variation of photoexcitation energy covering from low-energy absorption of CTX states to high-energy vibronic levels of donor $^1$Ex state, along with changes in other parameters, such as temperature, composition, and bias voltage. These works found remarkable robustness of IQE against input photon energy in a wide range of organic photovoltaic devices, suggesting that excess energy of a hot state is wasted via rapid internal conversion processes and is therefore not key for charge separation [12].

What has been persistently reported in the spectroscopy community, on the other hand, is the ultrafast generation of free charge carriers preceding internal conversion of intermediate states [15-17]. Reported timescales are often tens of femtoseconds, which cannot be captured by electrical characterizations in the steady-state, and vary considerably with the pump energy. For example, a study on the PCPDTBT/PC$_{60}$BM blend, reported by Grancini et al., showed that the generation rate of charge separated states (CS; free polarons) is twice faster upon photoexcitation of higher-lying exciton state than that from the lowest-energy exciton state [15]. Authors suggested that the excess energy of the exciton state is directly projected into the high-energy CTX states, which have a higher degree of charge delocalization than the states in the lower energy. Due to their excess energy that aids barrier crossing, but also to the decreased Coulomb binding which reduces the barrier height as a result of delocalization, high-energy delocalized CTX states have been proposed as the origin of efficient free charge carrier generation.

Additional evidence came from pump-push-probe spectroscopy which allows tracking of charge separation in real-time. In a series of experiments conducted by Friend group, authors analyzed how electroabsorption (EA), generated by local electric field of electron-hole pairs, evolves in time to study the charge separation dynamics [4, 18]. In order to isolate EA from congested pump-probe spectra, near-infrared (NIR) push pulse was employed to selectively perturb the electron-hole distance. The push on-off difference signal at different pump-push delays provided a direct visualization of EA evolution, which could be translated into an electron-hole pair distance increasing on an ultrafast timescale and ultimately probed the dynamics of the delocalization.

The energy of the push employed in these experiments is typically in the range ~0.4-0.6 eV [19, 20], which is very large considering that the estimates of CTX binding energy are a few hundred millielectronvolts [21, 22]. In this regard, mid-infrared (MIR) photoexcitation with energy less than the CTX binding energy can decisively tell whether charge separation is aided by charge delocalization or is simply promoted by an energy input large enough to cross the potential barrier instantly. Also, in the MIR region, vibrational and electronic transitions of transient charged species, i.e. free polarons and CTXs, coexist. In small donor-bridge-acceptor triad systems, a push targeting bridge vibrations has been shown to increase or decrease the yield of charge transfer [23-25]. Thus, MIR has the potential to disentangle the role of vibrationally and electronically hot states on the charge

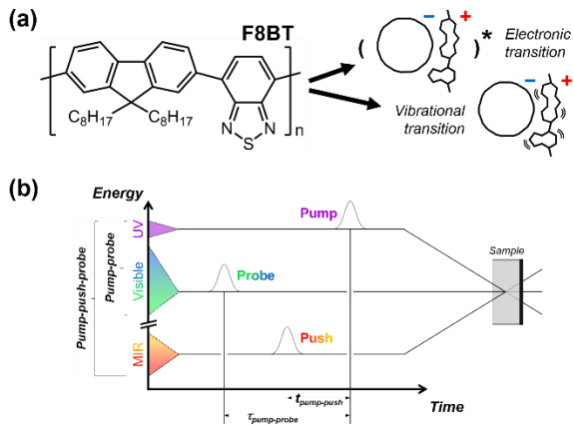

*Figure 1:* (a) Molecular structure of poly(9,9-dioctylfluorene-alt-benzothiadiazole) (F8BT) and the consequences of mid-infrared excitation of a charge transfer exciton generated at the F8BT:C60 interface. (b) Schematic of pump-push-probe spectroscopy employed in this study.

separation dynamics in the organic solar cell materials, which remains unexplored.

Here, we employ pump-push-probe spectroscopy to study the effect of a low-energy photoexcitation on the dynamics of a charge transfer exciton (CTX) in a polymer:fullerene bulk heterojunction material (Figure 1). Push energy was tuned from 0.12-0.25 eV in the mid-infrared (MIR) region to monitor how vibrational and low-energy electronic transitions of transient species respond to the given excess energy. To ensure that the energy of the push is smaller than the CTX binding energy, poly(9,9-dioctylfluorene-*alt*-benzothiadiazole) (F8BT), whose fullerene bulk heterojunction is known to have poor external quantum efficiency (EQE) [26] and power conversion efficiency (PCE) [27], was chosen.

We found three types of CTXs, delocalized, localized, and trapped, generated at the interface of F8BT:$C_{60}$ by analyzing the spectral signature found in both pump-probe (PP) and pump-push-probe (PPP) spectra. The "hot" CTXs created immediately after the ultrafast charge transfer dynamics following photoexcitation display hole polaron absorption in the visible region that subsequently redshifts, indicating that initially the CTX is delocalized among multiple polymer chains but rapidly localizes within a single chain. We show that a push pulse resonant with the low-energy charge transfer absorption at 0.19 eV can repopulate delocalized CTXs from localized CTXs that nevertheless retain a degree of intrachain hole delocalization, providing further chance to generate CS state. However, the push pulse was shown ineffective to promote delocalization at long pump-push delays from what we deem are trapped CTXs, where the polaron not only is localized within a single chain but also exhibits no intrachain delocalization. We claim that a degree of charge delocalization, either intra- or interchain arising from interchain coupling or regioregularity, is essential to allow charge transfer absorption and increase effective electron-hole separation, which then reduces the Coulomb barrier for charge separation.

**RESULTS**

We performed pump-probe (PP) and pump-push-probe (PPP) experiments on both a pristine F8BT and a blend F8BT:$C_{60}$ thin film, whose chemical preparation is described in the Supporting Information (SI). The pump-probe measurements, in combination with the steady-state absorption measurements of the two samples (Figure S1), serve the main purpose of identifying the states involved in the photoexcitation and establishing their spectral signature. This detailed knowledge of the photophysics of the systems will set then the basis for investigating the MIR-induced changes of the spectra by means of PPP spectroscopy.

**A. Pump-probe experiments**

Experimental details on the PP measurements are given in SI. The PP dynamics has been analyzed in both samples by using the Decay Associated Spectra (DAS) approach [28], which is a well-established tool in the non-equilibrium community to extract the transient absorption spectra by multiexponential decay fitting of the PP curve.

The spectra of F8BT pristine film have four global exponential time constants, which are shown as evolution associated spectra in Figure 2a. The spectra from $t_1$ through $t_3$ are characterized by the same spectral signatures. Furthermore, the presence of a quasi-isosbestic point at 1.85 eV indicates that the spectral evolution up to $t_3$ is dominated by a single process.

In order to identify the species involved in the decay, we focus on the initial spectrum with time constant $t_1$. It shows four distinct bands: two negative bands centered at 2.56 and 2.26 eV, one positive band at 1.33 eV, and a shoulder band at 1.66 eV. Based on the steady-state measurements (Figure S1), the negative bands at 2.56 and 2.26 eV correspond respectively to the lowest energy absorption and fluorescence peaks and are therefore ground-state bleach (GSB) and stimulated emission (SE). The photoinduced absorption at 1.33 eV has been studied by Stevens et al. and was assigned to the excited-state absorption (ESA) of singlet exciton ($^1$Ex) based on the exciton-exciton annihilation dynamics observed at high pump fluence [29]. As shown in Figure 2b, pump fluence-dependent decay dynamics are also reproduced in our sample, which revalidates the assignment of this band. This is also consistent with the study by Denis et al. of F8BT in solution which identified an ESA around 1.3 eV and a smaller ESA around 1.55 eV, both assigned to an excitation from the singlet exciton $^1$Ex [30]. ESA at 1.3 eV was found to arise from excitation to a higher exciton state, while the ESA at 1.55 eV was assigned to a charge transfer from a benzothiadiazole (BT) unit to another BT unit.

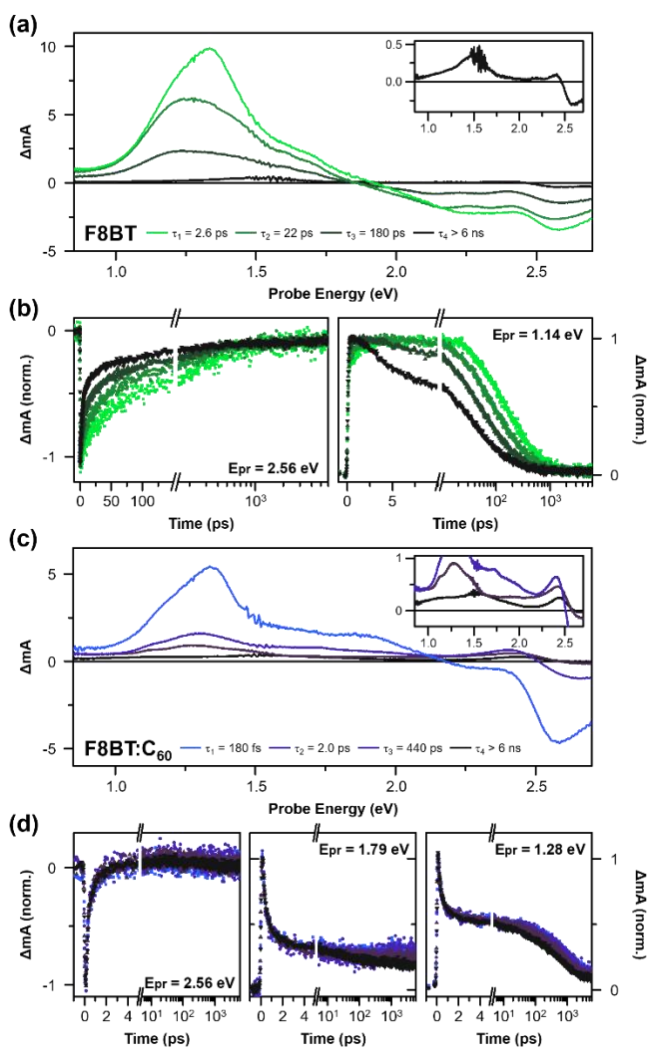

***Figure 2:*** *Evolution associated spectra and the representative decay traces of (a-b) F8BT pristine and (c-d) F8BT:$C_{60}$ blend films. Insets are scaled to show the spectra with small $\Delta A$ signals. Pump-probe (PP) spectra were obtained upon photoexcitation at 2.82 eV (440 nm) with pump fluence of 0.4, 0.8, 1.6, and 3.2 µJ cm$^{-2}$.*

As the spectra associated with the second and the third time constants do not develop a new band, they are also dominated by the decay of $^1$Ex to the ground state. We observe small red-shift in the ESA of $^1$Ex, accompanied by a red-shift in the SE, which originates from the exciton downhill migration to lower energy sites of the density of states, which is also reported in other pristine conjugated polymer films [31, 32]. The shoulder band at 1.66 eV also follows the decay profile of the 1.33 eV $^1$Ex ESA, suggesting that the main contributor of the peak is the same singlet exciton, consistent with the results in Refs [29, 30]. Yet, their peak ratio gradually changes (Figure S5), indicating that the physical origin of the two peaks may not be the same. In this respect, in the case of an archetype conjugated homopolymer, P3HT, initial PP spectra display a single broad Gaussian ESA band associated with singlet exciton, which decays and gives rise to polaron ESA at a different energy in the late time spectra. Similarly, the presence of this shoulder peak in our PP spectra may contain signal from a different state with some charge transfer character at later times. This will be confirmed by our PPP experiments.

The last spectrum associated with $t_4$ (Figure 2a, inset) is drastically different from the rest and is composed of a broad peak centered at 1.49 eV, GSB, and a small positive absorption at 2.39 eV. The band at 1.49 eV is reported as ESA of triplet exciton ($^3$Ex) by Lee et al., which shows up in the quasi-steady-state photoinduced absorption of a blend film of F8BT with iridium (III) complex capable of triplet sensitization [33]. The derivative-like shape near GSB (~2.5 eV) likely arises from electroabsorption (EA), which is the ground-state absorption red-shifted by local electric field of bound charge pair, that overlaps with GSB in the opposite sign [34]. This band is intensified when charge transfer yield increases, as it will become evident in the $C_{60}$ blend film.

In Figure 2c, we plot the decay spectra measured in the F8BT:$C_{60}$ blend film. The initial spectrum $t_1$ shows attenuated SE band and a broad ESA that appears as a hump near 1.9 eV. The $^1$Ex ESA at 1.33 eV manifests instead at the same energy as that in the pristine film. An attenuated SE in the initial spectrum suggests that a portion of the photoexcitation produces CTX within the pulse duration. The $^1$Ex ESA, along with the GSB, rapidly decays in $t_1 = 180$ fs to give the $t_2$ spectrum with prominent EA at 2.40 eV and completely quenched SE. Since the kinetics is independent of pump fluence and EA develops in the following spectrum, this process can be associated with charge transfer from F8BT to a proximal $C_{60}$. Surprisingly, in the $t_2$ spectrum, ESA at 1.30 eV remains while SE is completely quenched. This ESA persists in $t_3$ and survives until $t_4$ spectrum which contributes to an increased signal in the red side of the $^3$Ex ESA when compared to the $t_4$ spectrum of the pristine film. Also $t_4$ spectrum contains additional signal in the blue side of the $^3$Ex ESA band, which persists from the earlier spectra. The hump ESA around 1.9 eV in $t_1$ becomes a dragging shoulder spanning 1.5-2.0 eV in $t_2$. This decays in intensity in $t_3$ but remains stagnant until $t_4$.

In order to emphasize the role of the acceptor and extract the spectral evolution of the early photoexcited states in the $C_{60}$ heterojunction, we subtracted the $t_1$ spectrum of the pristine film from the $t_1$ spectrum of the blend film, which leaves this new ESA at 1.94 eV associated with the time constant $t_1$ (Figure 3a). For the time constant $t_2$, we subtracted the $t_3$ spectrum from the $t_2$ spectrum, both from the blend film (Figure 3b). The extracted band associated with $t_2$ spectrum shows a maximum at 1.72 eV. Peak position of these new ESAs do not further red-shift until $t_4$ (Figure S6). We attribute the 1.72 eV ESA to the F8BT$^{\bullet+}$ polaron absorption which matches the reported peak position of F8BT$^{\bullet+}$ absorption from an injected hole following the assignment by Bird et al [35].

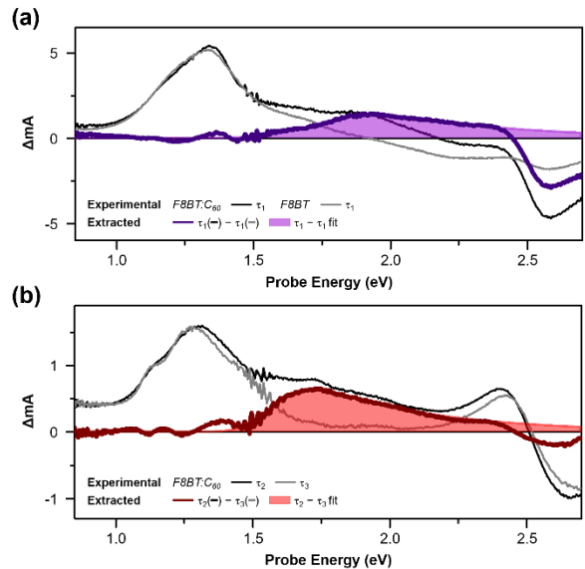

*Figure 3*: Polaron absorption spectra extracted from (a) $\tau_1$ and (b) $\tau_2$ pump-probe spectra of F8BT:$C_{60}$ blend film. Polaron absorption in $\tau_1$ spectrum (purple solid line) was obtained by subtracting $\tau_1$ spectrum of the pristine film from $\tau_1$ spectrum of the blend film (grey and black solid lines in panel a, respectively). Polaron absorption in $\tau_2$ spectrum (maroon solid line) was obtained by subtracting $\tau_3$ spectrum from $\tau_2$ spectrum both from the blend film (grey and black solid lines in panel b, respectively). Polaron absorption peaks are fitted with exponentially modified Gaussian functions to determine their peak positions.

We claim that the higher energy peak position in $t_1$ is a result of a greater degree of hole delocalization of a "hot" delocalized CTX (DCTX) which retains an amount of excess energy with respect to the CTX with hole localized on a single chain. The lower energy absorption in $t_2$, which matches instead with the reported peak position of F8BT$^{\bullet+}$ absorption from an injected hole [35], signifies that it originates from the localized hole of a localized CTX. A large difference in the polaron absorption peak position in the visible wavelength has been reported in other conjugated polymers, e.g., P3HT, and was explained with an electronic structure model constructed from a coupled chain aggregate, which predicts that delocalization will blueshift absorption with respect to that of a localized hole [36].

From here, we will refer to the peaks in Figure 3a and 3b as DP2 and P2, to indicate that these absorptions originate from higher-energy transition of delocalized and localized polaron, respectively. Absence of pump fluence-dependence in the time constant $t_1$ indicates that it is a unimolecular process. Since $t_1$ dynamics greatly reduces the intensity of GSB by 79%, we suggest that the "hot" DCTX is mostly quenched by geminate charge recombination. What is left after this process is the absorption at 1.72 eV. This suggests that the remaining population becomes localized CTX. More in particular, their long lifetime suggests that these are trapped CTX (TCTX), which will be confirmed by the PPP measurements.

Transition between $t_2$ and $t_3$ spectrum shows a large reduction in the F8BT$^{\bullet+}$ polaron absorption (Figure 2c inset), accompanied by the decay in both ESA around 1.30 eV and the GSB. We attribute this process to, again, charge recombination. Interestingly, both $t_2$ and $t_3$ spectra show ESA at the energy of $^1$Ex ESA ~1.30 eV even when SE has completely vanished and DP2 or P2 is present. This ESA also decays with the polaron absorption at the charge recombination time $t_2$. It is tempting to assign such coexisting spectral signatures to branching dynamics due to inefficient CT. But, because the intensity of the ~1.30 eV $^1$Ex ESA does not properly mirror the complete quenching of SE in the later spectra, the origin of this ESA at later times is a different state from $^1$Ex. Similar behavior had been a source of debate in other conjugated polymers [37-39], until Wang and Psiachos showed that the origin of the mixed signal is ascribable to a CTX [40, 41]. While initially most absorption comes from the singlet exciton $^1$Ex, at later times, it is the CTX absorption that dominates. Being a CTX a superposition of a neutral exciton and a polaron pair in presence of a significant intermolecular coupling, configuration interaction from CTX allows separate excitation of constituent excitonic and polaronic states that contributes to their own spectral signatures. Importantly, ESA from the CTX is expected at the energy of the $^1$Ex ESA. Thus, we finally attribute simultaneous manifestation of $^1$Ex ESA and DP2/P2 to the spectral signature of CTX. Charge separated (CS) states share a minor contribution to the spectra, as shown in the PPP experiment in the following.

The last $t_4$ spectrum again shows dominant $^3$Ex ESA at 1.49 eV, but the red and blue side of the peak contain additional ESA, which presumably originate from long-lived TCTX. This is also supported by the EA signature, which is more intense than the $t_4$ spectrum of the pristine film.

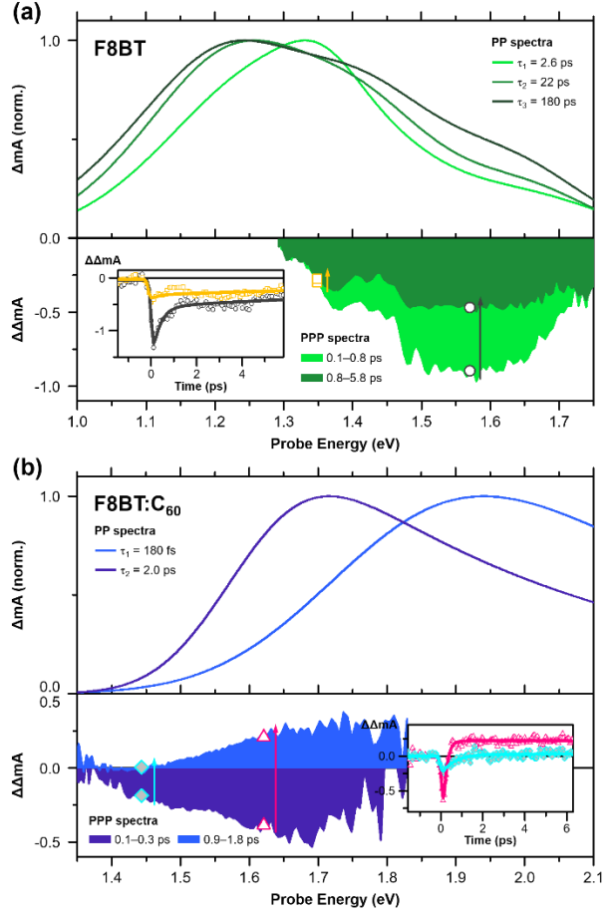

**Figure 4:** Pump-push-probe experiment results on (a) F8BT pristine and (b) F8BT:C$_{60}$ blend film. Push-induced difference spectra (bottom panels; integrated over the push-probe time delay t$_2$ indicated in each legend) and the pump-probe spectra (top panels; evolution associated spectra analyzed with time constants indicated in each legend) are plotted together for comparison. The pump-probe spectra from respective time constants are analyzed with multi-Gaussian peak functions to present only the excited state absorption of singlet excitons in (a) and polaron absorption in (b) (see Supporting Information for more details). Insets show kinetics of the push-induced difference signal at probe wavelengths indicated by the symbols in the spectra (○, 1.57 eV; □, 1.35 eV; △, 1.62 eV; ◇, 1.44 eV) and adjacent arrows point the direction of changes over time. Push-induced difference spectra shown here are obtained upon photoexcitation at 2.76 eV (450 nm) with pump and push fluence of 25 and 450 μJ cm$^{-2}$, respectively.

## B. Pump-push-probe experiments

Input of sub-bandgap energy has been reported to change the electron-hole distance and increase the charge separation yield [19, 20]. In the mid-infrared (MIR) region two different absorbers can contribute: (1) vibrational motion in the charge transfer coordinate and (2) electronic transition of the charged polaronic states. To test if MIR excitation can contribute to creating free charge carriers and what is the mechanism behind it, we added a MIR push pulse in the pump-probe pulse sequence and monitored the spectral and kinetic changes induced by it. We highlight that the MIR photon energy is sufficiently low that, except for a fast optical Stark effect, no push-probe signal is present in the absence of the pump (Figure S7). Information on the PPP setup is given in Ref. [42] and further experimental details are discussed in the SI.

We first investigate the effect of a push energy of 0.18 eV on the F8BT pristine film. At pump-push time delay t$_1$ = 0 ps, changes in the pump-probe signal ($\Delta\Delta A = \Delta A^{push|ON} - \Delta A^{push|OFF}$, Figure 4a bottom panel) are persistently negative in all push-probe time delays t$_2$ > 0 ps, which indicates that the population of the excited-state species associated with the pump-probe signal decreases with the push pulse. Increasing $t_1$ neither changes the sign nor the spectral shape of the $\Delta A$ (Figure S8). Two bands in the $\Delta\Delta A$ signal are centered at approximately 1.58 and 1.37 eV (Figure S11) and they roughly coincide to the peak positions of the shoulder and the main peak of the $^1$Ex ESA, respectively, as shown in the top panel. However, the peak intensity ratio is reversed in the $\Delta\Delta A$ spectra. Notice also that the decay of the higher energy band has prominent decay fast component of 350 fs that is almost absent in the lower energy band (Figure 4a bottom panel inset). This shows that the states that give rise to both ESAs respond differently to the push pulse, which is at odds with the possibility that they originate from the same singlet exciton state $^1$Ex, further supporting the idea that not just a single state but two, the exciton and the CTX, are contributing to the ESAs even immediately after photoexcitation.

In the F8BT:C$_{60}$ blend film, at $t_1$ = 0 ps, the $\Delta\Delta A$ signal is negative overall at small push-probe delay $t_2$, but turns into positive at $t_2$ > 400 fs, which persists (Figure 4b bottom panel). This means that when the push acts on the system it depletes the states that give rise to the transitions in the pump-probe spectra, but its effect on the long run (> 400 fs) is to enhance the population of such states. The negative spectrum at $t_2$ < 400 fs has a maximum at around 1.67 eV and the positive spectrum that follows is much blue-shifted. When compared with DP2 and P2 peaks at 1.94 and 1.72 eV from $t_1$ and $t_2$ PP spectra, which are from delocalized and localized CTX, respectively, it becomes clear that the push pulse effectively increases delocalized states more than localized CTXs. The rise time is 160 fs at probe energy 1.62 eV (pink trace in the inset of Figure 4b). Interestingly, the behavior of $\Delta\Delta A$ signal changes drastically at different pump-push delays. At $t_1 \geq 1$ ps, $\Delta\Delta A$ displays only negative spectra that are centered towards DP2 and the sign does not change throughout the $t_2$ window (Figure S9). Dependence on the pump-push delay $t_1$ indicates that the push is acting on different excited-state species; because DCTX gets trapped within $t_1$ = 180 fs, at pump-push delay of $t_1 \geq 1$ ps only localized CTX or the other product state accessible from DCTX, are expected to be populated.

Low energy charge transfer absorption from polarons in the 0.1 eV range has been interpreted in the past as interchain charge transfer from delocalized polarons in polymers where significant interchain interactions exists. More recently, Spano and coworkers showed that this kind of transitions are also expected in single polymer chains where the polaron delocalizes within the chain, with the strength of the transition being directly proportional to the polaron size [43-46]. According to these studies, the push can only induce transitions in states that exhibit either interchain or intrachain delocalization.

Given that the negative band develops around the DCTX absorption, implying that localized CTX populated at $t_1 \geq 1$ ps do not respond to the push pulse, our results suggest that the CTXs exhibit no inter- or intrachain delocalization, which is consistent with our assignment of these as trapped CTX based on the pump-probe measurements.

The nature of the charge transfer absorption that gives rise to the band at DCTX absorption will be further discussed in the Discussion section. It should also be stressed that multiphoton absorption effects are unlikely to occur at the push fluences employed here. To rule out this possibility, we performed pump-push-probe experiments with varying push fluences (Figure S10) that confirm that the differential signals in Figure 4 scale linearly with the fluence of the push pulse.

To better understand the origin of push-sensitive excited-state species, we tuned the push energy in the range of 960-2100 cm$^{-1}$ (0.12-0.26 eV), which covers the vibrational spectrum of F8BT as shown in Figure 5b. Push-energy dependent $\Delta\Delta A$ response at $t_1$ = 0 ps are plotted in the upper and lower panel of Figure 5a for F8BT pristine and F8BT:C$_{60}$ blend films, respectively. In the F8BT pristine film, $\Delta\Delta A$ shows two different curves as a function of probe energy (Figure 5a upper panel). $\Delta\Delta A$ at 1.33 eV shows a strong negative signal only when the push pulse is tuned to 1430 cm$^{-1}$, which coincides with the strong infrared (IR) absorption by C=C stretching mode, whereas $\Delta\Delta A$ at 1.57 eV displays a negative response over a broad range of push energies. This remarkably different response to the push excitation further supports the idea that the two states have different physical origin. In particular, the broad absorption in the MIR at 1.57 eV is a characteristic of polaron absorption, which is also observed in other conjugated polymers. As such, we assign the band centered at 0.19 eV to the mid-infrared charge transfer (CT) absorption, which, as discussed above, can have an interchain or intrachain character as long as the polaron is delocalized within a chain. The response of the peak at 1.57 eV by CT absorption supports our assignment of the CTX contribution to the shoulder ESA. In the same line, response at 1.33 eV associated with the vibrational spectrum of neutral F8BT reinforces the assignment of the corresponding band as ESA from singlet exciton. It is evident that both electronic excitation of CTX and IR-active vibrational excitation deplete the respective states in the pristine F8BT film.

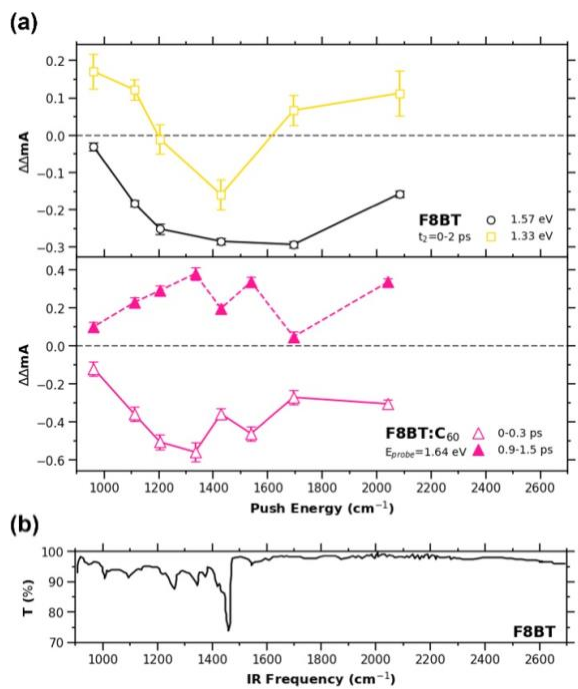

***Figure 5:*** *(a) Push-induced difference signal as a function of push energy for F8BT pristine and F8BT:$C_{60}$ blend films (upper and lower panels, respectively). For F8BT pristine film, the difference signal monitored at 1.57 and 1.33 eV (grey and yellow lines) shows different push-energy-dependence. For F8BT:$C_{60}$ blend film, spectra before and after (solid and dotted lines, respectively) the sign change of the difference signal monitored at 1.62 eV are plotted. The error bars indicate the standard deviation calculated – for a given probe and push energy – over a fixed interval of negative delays of each time trace and averaged over the delay interval indicated in the panels. (b) Steady-state infrared transmittance spectrum of F8BT.*

$\Delta\Delta A$ of F8BT:$C_{60}$ blend film again shows a broad absorption over the MIR range tested, but at two push energies, 1430 and 1700 cm$^{-1}$, the response becomes attenuated (Figure 5a lower panel). The signal is negative before $t_2$ = 400 fs and turns into positive afterward, as shown in Figure 4b, without altering the push-energy-dependent spectral shape. Again, the broad absorption can be attributed to CT absorption from the CTX, yet the difference in the blend film is that the electronic transition caused by push eventually enhances ESA from CTX. One would be tempted to attribute the dip at 1430 cm$^{-1}$ to the same vibrational excitation of the neutral species, but the appearance of another dip at 1700 cm$^{-1}$ suggests that the two peaks originate from the vibrational modes of the F8BT$^{\bullet+}$ in the CTX, which is supported by DFT calculations (Figure S12). Overall, the data show that while electronic CT absorption can eventually enhance CTX states, vibrational excitation tends to deplete these states in the polymer:fullerene bulk heterojunction material. All excited-state species identified here are summarized in Figure 6 into a schematic energy diagram with associated spectral signatures.

**DISCUSSION**

CTXs are influenced by parameters such as donor-acceptor distance, orientation, and dielectric environment [47, 48]. In this respect, owing to conformational heterogeneity at the donor-acceptor interface in the polymer bulk heterojunction we expect a distribution of CTXs. In this work, we have identified three distinct classes of CTXs: delocalized (DCTX), localized (CTX), and trapped (TCTX) charge transfer excitons. These species appear to differ in their energies, lifetimes, and most importantly, response to the push input. In what follows, we elaborate on the dynamic processes that produce these states and how PPP spectra, which seem irreconcilable with the bulk dynamics from PP spectra, can elucidate their nature and varying extent of polaron delocalization.

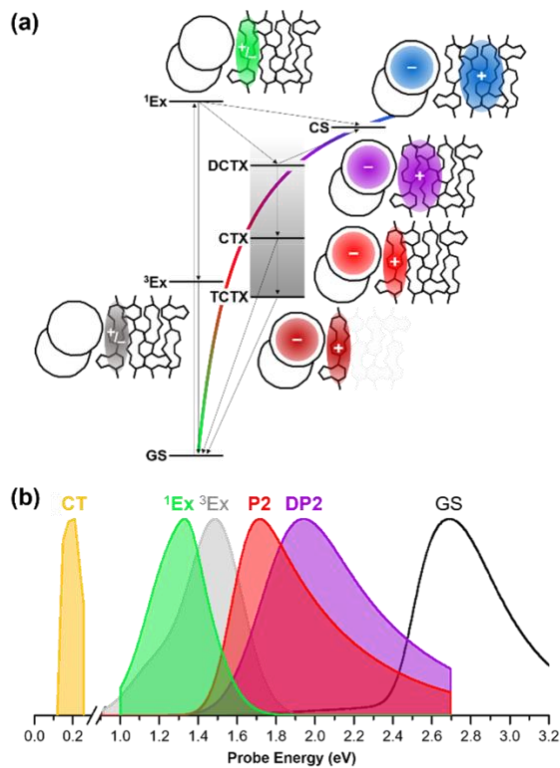

*Figure 6:* (a) Energy diagram of all excited states (GS: ground state, $^1$Ex: singlet exciton, $^3$Ex: triplet exciton, CTX: localized charge transfer exciton, DCTX: delocalized charge transfer exciton, TCTX: trapped charge transfer exciton, CS: charge separated state) identified using pump-probe and pump-push-probe experiments. Spatial distribution of excitation is depicted next to each state. Thin arrows indicate possible transition pathways identified in this work. (b) Absorption spectra from all excited-states extracted from pump-probe and pump-push-probe experiments. (DP2: higher-energy transition of delocalized polaron, P2: higher-energy transition of localized polaron)

The DCTX is a hot CTX state that retains excess energy immediately after charge separation from $^1$Ex. We assign DCTX states to lie closer to the dissociation threshold of the qualitative Coulomb potential (Figure 6a), implying that the mean electron-hole (e-h) distance is large. The CTX species, on the other hand, is assigned to be further from the dissociation threshold, and to exhibit a shorter mean e-h distance, yet it is thermally unequilibrated like DCTX.

The electronic configuration of the DCTX and CTX is further clarified in Figure 7a, where the superposition of a neutral exciton and a bound polaron pair is depicted for the simplest case of a two-site electron donor and single site acceptor. The degree of polaron delocalization influences the energies of the allowed optical transitions, as has been discussed by Ref. [36]. In particular, when a polaron resides in a single chain (CTX) (top), two transitions, P1 and P2, appear typically in the infrared and visible region, respectively. But when a polaron is delocalized (DCTX), such as when the hole delocalizes over two chains (bottom), doublets of states formed via interchain coupling of a charged polaron and a neutral ground state alters the allowed transitions to DP1 and DP2, which are red- and blue-shifted with respect to P1 and P2, respectively. As a consequence, DCTX is characterized by the DP2 transition, while CTX by the P2 transition, which lies at lower energy. This is the basis of our assignment of the 1.94 eV band (Figure 3a) as delocalized F8BT$^{•+}$ polaron ESA and the 1.72 eV (Figure 3b) as localized F8BT$^{•+}$ polaron ESA as previously reported in the literature [35].

In the F8BT:C$_{60}$ blend, the dominant species at early pump-probe delay is DCTX, as pronounced DP2 transition of delocalized polaron indicates (Figure 3a, $t_1$). Surprisingly, the PPP spectrum at pump-push delay $t_1 = 0$ initially shows a negative P2 band of localized polarons at $t_2 < 400$ fs (Figure 4b). At pump-push delays $t_1 \geq 1$ ps, PPP spectra show negative $\Delta\Delta A$ that matches DP2 transition (Figure S9). However, in the corresponding PP spectra represented by $t_2$, P2 transition is prominent (Figure 3b). In other words, at short pump-push delays, when the system is expected to exhibit prominent DP2 absorption, the push is reducing mostly the P2 ESA. This suggests that 1) CTX are generated along with DCTX during photoexcitation and that 2) CTXs carry higher oscillator strength than the prevailing DCTX at the given push energy. Instead, at long times, when localization dominates, the push is mostly reducing absorption in the DP2 region. This in turn suggests that 1) the origin of DP2 transition in the PPP spectrum is no longer the DCTX that displays significant PP signal at $t_1 = 0$ and 2) DP2 band in the PPP spectra is produced by a state with polaron character resembling DCTX.

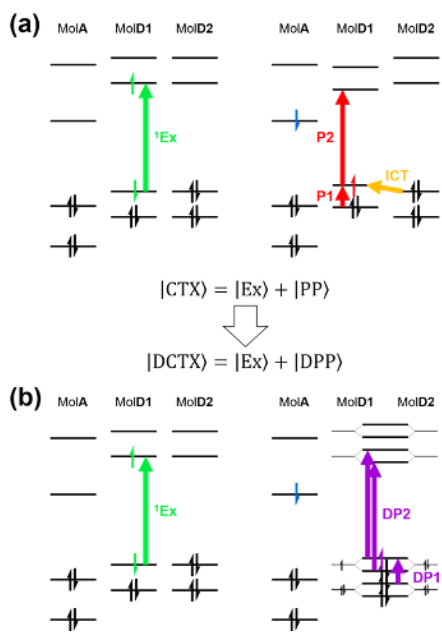

*Figure 7:* Electronic configurations of (a) charge transfer exciton ($|CTX\rangle$) and (b) delocalized charge transfer exciton ($|DCTX\rangle$). A colored vertical arrow represents the most dominant configuration of a transition that corresponds to the excited-state absorption from excitonic ($|Ex\rangle$; 1Ex) or polaritonic ($|PP\rangle$; P1 and P2, $|DPP\rangle$; DP1 and DP2) component of a charge transfer exciton. The interchain charge transfer transition (ICT in yellow arrow) perturbatively mixes singly-charged donor (MolD1) and neutral donor (MolD2) into delocalized polaron, which displays excited-state absorption different from that of polaron before mixing.

To interpret these seemingly contradicting data from PP and PPP experiments, we turn again to the low-energy CT MIR band expected from polarons exhibiting delocalization. On the one hand, in systems with strong interchain interactions, the push pulse may induce interchain charge transfer (ICT) transition, where charge transfer happens from a delocalized polaron to higher excited delocalized polaron. Spano and co-workers have shown that these bands are also expected in single chain polymers as long as there is polaron delocalization within a chain [43-46]. A push pulse can then induce transitions on either DCTX or localized, but not trapped CTXs. Therefore, at short pump-push delays, the push induces transitions in localized CTX states and to a lesser degree in DCTXs, initially depleting them, as the negative $\Delta\Delta A$ suggests (purple curve in Figure 4b bottom panel). The difference spectra then turns positive, suggesting that relaxation dynamics from the push excited CT states enhance the population of DCTX states after 400 fs. Instead, at long pump-push delays $t_1 \geq 1$ ps, the pump acts on TCTX and CS states. Since TCTXs show P2 ESA, the prominent negative DP2 band suggests that the push is acting on the other available states, namely the long-lived CS states. We believe that the push induces CT transitions in the (possibly delocalized) CS state, and that its effect is to deplete this state by inducing charge transfer that can either further separate the electron or hole (without changes in the spectra) or reduce their distance leading to increased charge recombination.

Finally, the results of our PPP experiment contrast with other studies where push promotes delocalization and eventually charge separation at pump-push delays much larger than the charge transfer time [19, 20]. It should be stressed, however, that the energy of the push pulse used in studies reported in the current literature is ~0.4-

0.6 eV, which is larger than the activation energy of trapped singlet excitons and polarons [49, 50]. It is also comparable, or often larger than, the CTX binding energy estimated from lowest-lying CTX [21, 22]. Those studies may therefore not be sensitive to the fleeting presence of hot states that appear to favor delocalization.

**CONCLUSIONS**

In an organic bulk heterojunction material comprising F8BT and $C_{60}$, we have demonstrated that the hole delocalization of CTX is enhanced by a push pulse that targets the low-energy charge transfer absorption centered around 0.19 eV. Our main evidence of the delocalized character of the charge transfer exciton DCTX compared to the well-known charge transfer exciton CTX was the blueshift in the peak position of the polaron absorption in the visible region as a result of interchain coupling.

Interestingly, we were able to identify two kinds of states that were predominantly sensitive to the push: CTXs localized in a single chain, which nevertheless retain a degree of intrachain delocalization, and charge separated states CS. The push acting on the localized CTX eventually enhanced the population of DCTX on an ultrafast timescale, which is expected to increase CS due to the increased electron-hole separation. Instead, the push contributed to charge recombination of the CS states. Since CTXs are expected to be sensitive to the MIR push only when intra- or inter-chain hole delocalization is present, we claim that when interchain coupling is weak or intrachain delocalization is not supported, trapped CTX (TCTX) will dominate, and the excess energy is wasted.

Vibrational excitation on the same energy scale, on the other hand, was shown to deplete the CTX as well as $^1$Ex population, likely due to quenching. This adverse effect suggests that the vibrationally hot state, unless it is coupled to a charge separated state directly, can distribute its energy to other vibrational coordinates that couple to dissipative pathways, including those that promote charge recombination.

Our results emphasize the importance of engineering both the energy offset between the singlet exciton and charge transfer states and local polymer morphology to optimize high-yield charge separation. They also show the strength of the pump-MIR push-probe technique that we have used in shedding light beyond what is readily available in pump-probe experiments into the physical nature and dynamics of the states involved in the ultrafast photophysics of polymer:fullerene blends.

**SUPPORTING INFORMATION**

Additional experimental and computational details, additional data including: steady-state absorption and fluorescence of F8BT and F8BT:$C_{60}$ in various solutions and thin films, evolution associated spectra and decay traces of the pristine F8BT film, excited-state absorption spectra of the photogenerated polaron in F8BT:$C_{60}$, push-probe maps, ΔΔA spectra at different pump-push delays, ΔΔA spectra at different push fluences, steady-state infrared absorption spectrum of neutral F8BT, calculated extinction coefficients for neutral and radical cation F8BT trimers. (PDF)

**ACKNOWLEDGEMENTS**

Financial support was provided by the Division of Chemical Sciences, Geosciences and Biosciences, Office of Basic Energy Sciences, of the US Department of Energy through grant no. DE-SC0015429. F.F. acknowledges financial support from the European Union's H2020 Marie Skłodowska-Curie actions (Grant Agreement No. 799408). D.F. was supported by the European Commission through the European Research Council (Project INCEPT, Grant 677488).

# Supporting Information

# Manipulation of Charge Delocalization in a Bulk Heterojunction Material Using Infrared Push Pulse

Angela Montanaro[1,2,•], Kyu Hyung Park[3,•], Francesca Fassioli[2,3,4], Francesca Giusti[1], Daniele Fausti[*,1,2], and Gregory D. Scholes[*,3]

[1] Department of Physics, University of Trieste, Via A. Valerio 2, 34127 Trieste, Italy; Elettra-Sincrotrone Trieste S.C.p.A. Strada Statale 14 - km 163.5 in AREA Science Park 34149 Basovizza, Trieste, Italy

[2] Department of Physics, University of Erlangen-Nürnberg, 91058 Erlangen, Germany

[3] Department of Chemistry, Princeton University, Princeton, New Jersey 08544, United States

[4] SISSA − Scuola Internazionale Superiore di Studi Avanzati, Trieste 34136, Italy

•*These authors contributed equally.*

AUTHOR INFORMATION

**Corresponding Author: *gscholes@princeton.edu; *daniele.fausti@elettra.eu**

**Table of Contents**

# 1. Experimental details

## 1.1. Chemicals and sample preparation

Poly(9,9-dioctylfluorene-alt-benzothiadiazole) (F8BT) ($M_n \leq 25{,}000$ Da) and fullerene ($C_{60}$) (99.5%) were purchased from Sigma-Aldrich and used without further purification. F8BT and $C_{60}$ were separately dissolved in chloroform (HPLC grade, $\geq 99\%$, contains amylene as stabilizer) by stirring 24 h, which were then mixed in equal weight ratio to prepare the blend film or diluted with the solvent to prepare the pristine F8BT film. Solutions were filtered out to remove any undissolved particles prior to spin coating. Fused quartz slides ($2.54 \times 2.54$ cm$^2$) were cleaned by sonication in detergent solution (2 vol%, Hellmanex), deionized water (Milli-Q), acetone (HPLC grade, $\geq 99\%$), isopropanol (ACS reagent grade, $\geq 99.5\%$), dichloromethane (ACS reagent grade, $\geq 99.5\%$, contains amylene as stabilizer) for 20 min. each and, finally, by oxygen plasma treatment for 5 min.

## 1.2. Steady-state spectroscopy

Steady-state absorption and fluorescence spectra of the pristine F8BT and F8BT:$C_{60}$ blend films, as well as F8BT solutions, were measured using Cary 100 UV-Visible spectrophotometer (Varian) and PTI QuantaMaster 400 (Horiba), respectively. Concentration of the F8BT solution samples were set to have absorbance $< 0.5$ to avoid the effect of self-absorption and self-quenching. Film samples tilted ~60° with respect to the incident excitation to direct emission to the detector in L-format, while avoiding collection of the reflected scatter.

## 1.3. Transient absorption spectroscopy

Pump-probe and pump-push-probe measurements were performed using two different experimental setups. The details of both setups are separately given in the following subsections.

1.3.1. Pump-probe spectroscopy

Description of the pump-probe setup is provided elsewhere.[1] The laser source is a Ti:Sapphire laser (Coherent Libra). Pump is tuned to 450 nm to resonate with the lowest-energy absorption band of F8BT for all samples discussed. The diameter of the pump and probe beam spots at the film or cuvette surface were 2 and 1 mm, respectively, which have been deliberately enlarged by placing the sample in front of the focal point. The pump intensity was varied systematically to test the dependence of excited-state dynamics on the pump fluence in the range of 0.4-3.2 µJ cm$^{-2}$. All measurements were performed at magic-angle configuration.

1.3.2. Pump-push-probe spectroscopy

Description of the pump-push-probe setup is provided elsewhere.[2] The light source employed is the Pharos laser by Light Conversion. Tunable mid-infrared (MIR) push pulse (70-250 meV) is generated by difference frequency generation in a GaSe crystal of two phase-locked near-infrared (NIR) pulses from a twin optical parametric amplifier (Orpheus TWIN, Light Conversion). Pump pulse is generated by second harmonic generation of a 20fs-long NIR output (900 nm) from non-collinear optical parametric amplifier (Orpheus-N, Light Conversion). A small portion (~1 µJ/pp) of the NIR output is used to generate the white-light (WL) probe by self-phase modulation in a sapphire crystal. Transmitted probe pulses are collected and dispersed by a blazed grating on a linear array of Si photodiodes (NMOS linear image sensor, Hamamatsu), whose pulse-by-pulse acquisition is synchronized with the laser repetition rate (5 kHz). Pump and push fluences were ~15 and ~500 µJ cm$^{-2}$, respectively. The implementation of a chopping system consisting of two optical blades synchronously running at known rotational frequencies (40 Hz for the pump and 20 Hz for the push) allows to separately acquire the dynamical response of the sample to: i) just the visible pump; ii) just the MIR push; iii) the joint response of both the pump and push. Direct subtraction enables the calculation of the MIR-induced change in the pump-probe signal ($\Delta\Delta A = \Delta A^{push|ON} - \Delta A^{push|OFF}$). To prevent photodegradation, nitrogen was blown on the sample for the entire duration of the experiment.

**1.4. Peak fitting procedure**

Large inhomogeneous broadening of polymer systems makes Gaussian peak fitting a desirable method of capturing the time evolution of the peak position and intensity, but other functions, e.g., log-normal function, are also employed to fit asymmetric peaks. We used a combination of Gaussian (for $^1$Ex, $^3$Ex) and exponentially modified Gaussian functions[3] (for P2 and DP2) to analyze the pump-probe spectra in the visible and near infrared region. Peaks are assigned according to the time evolution characteristic to the identified excited-state species, also by carefully reviewing the literature, as detailed in the main text. Decomposed peaks are used to plot Figure 3, 4, and 6, which show one-to-one correspondence of the photogenerated species to their spectral signatures.

## 2. Computational details

All calculations are performed with a model oligomer consisting of four fluorene (F1) units and three benzothiadiazole (BT) units connected in an alternating fashion (F1BT)$_3$-F1 (hereafter referred to as trimer). Terminals of the trimer are capped with methyl groups and dioctyl groups in the fluorene unit of the original polymer are substituted with dimethyl groups to reduce the cost of computation. Ground-state geometry optimization, frequency calculation, vertical transition energy calculation, and natural transition orbital (NTO) simulation are performed at CAM-B3LYP/6-31G(d) level of theory using Gaussian(R) 16 software.[4] Optimized structures of neutral and radical cation trimers are shown in Figure S13 and S14.

## 3. Supplementary Figures

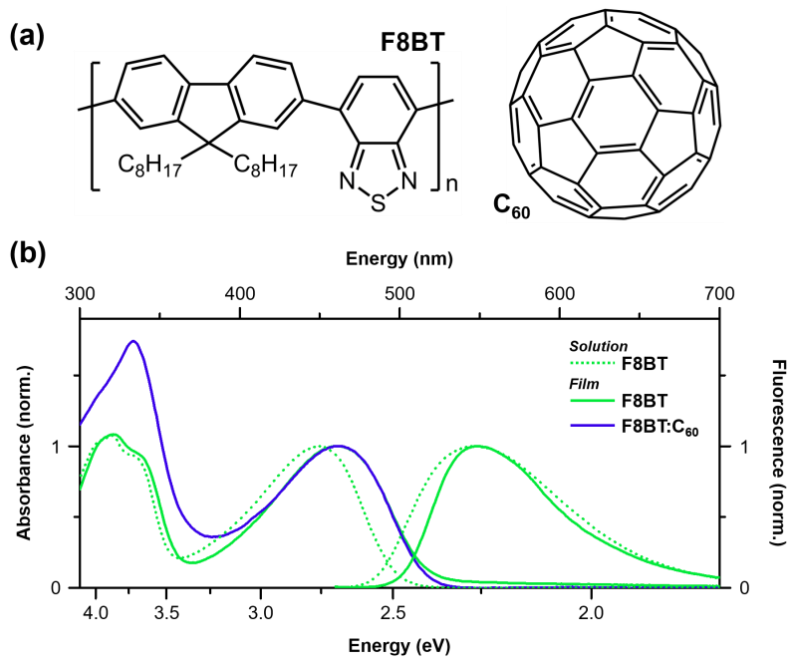

**Figure S1.** (a) Molecular structure of poly(9,9-dioctylfluorene-*alt*-benzothiadiazole) (F8BT) and fullerene ($C_{60}$). (b) Steady-state absorption and fluorescence spectra of F8BT in chloroform solution (green dotted lines), and F8BT pristine and F8BT:$C_{60}$ blend films (green and blue solid lines, respectively).

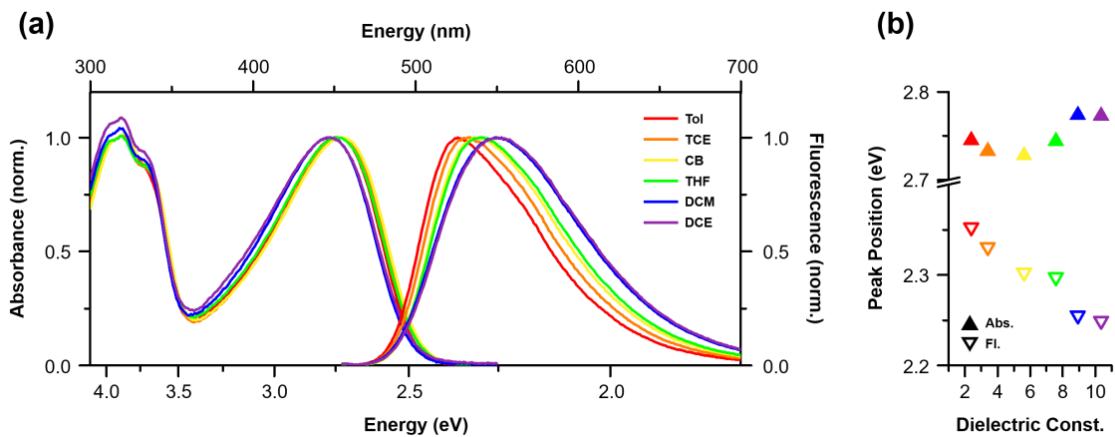

**Figure S2.** (a) Steady-state absorption and fluorescence spectra of F8BT in various solutions (Tol: toluene, TCE: trichloroethylene, CB: chlorobenzene, THF: tetrahydrofuran, DCM: dichloromethane, DCE: 1,2-dichloroethane). (b) Maxima of steady-state absorption (filled upward triangle) and fluorescence (empty downward triangle) bands as a function of dielectric constant of the solvent used.

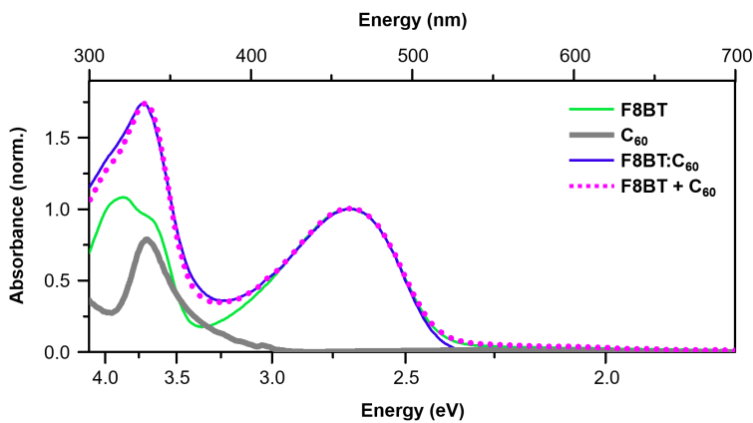

**Figure S3.** Steady-state absorption spectra of pristine F8BT (green solid line) and $C_{60}$ blend (blue solid line) films. Simulated steady-state absorption spectrum of the blend film (pink dotted line) is obtained by adding the steady-state absorption spectra of the pristine film (green solid line) and $C_{60}$ chloroform solution (grey solid line).

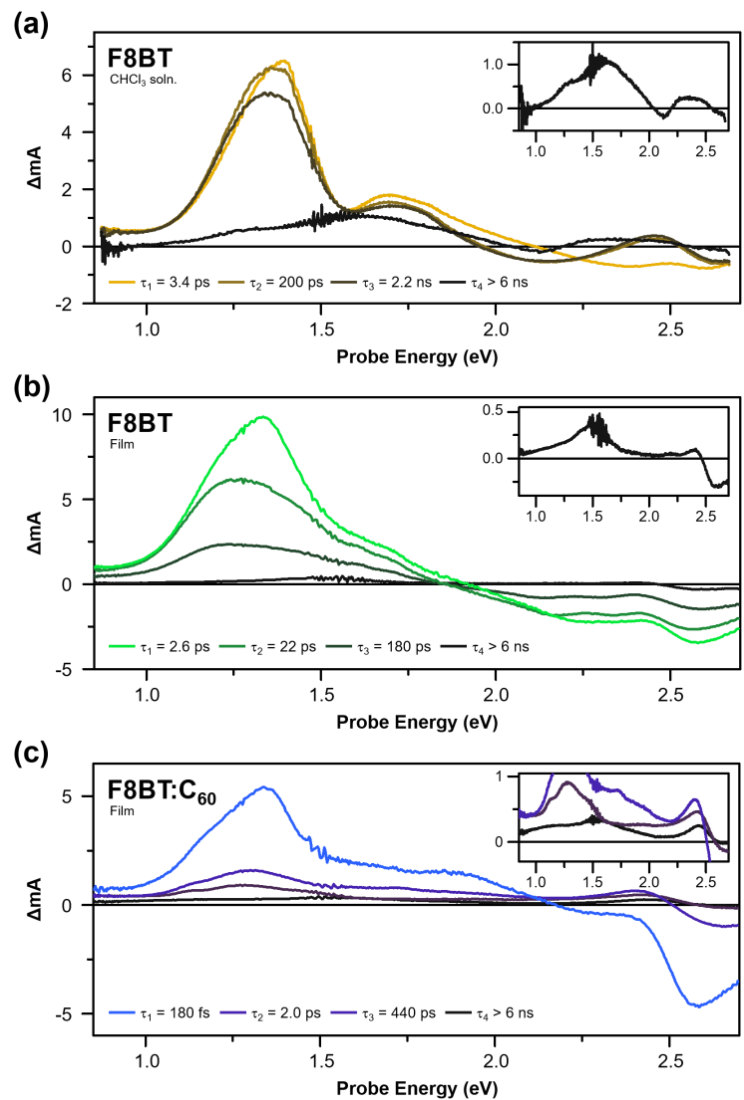

**Figure S4.** Evolution associated spectra of (a) F8BT in chloroform solution, (b) F8BT pristine film, and (c) F8BT:$C_{60}$ blend film. Insets are scaled to show the spectra with small ΔA signals.

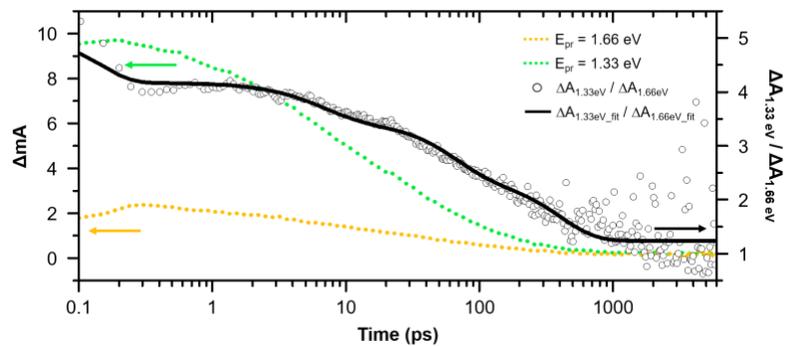

**Figure S5.** Decay traces of F8BT pristine film at probe energy 1.66 and 1.33 eV (yellow and green dotted line, respectively). These decay traces are analyzed with 4-exponential functions and the ratio of fitted curves (black solid line) are superimposed to the same plot from the raw data (○).

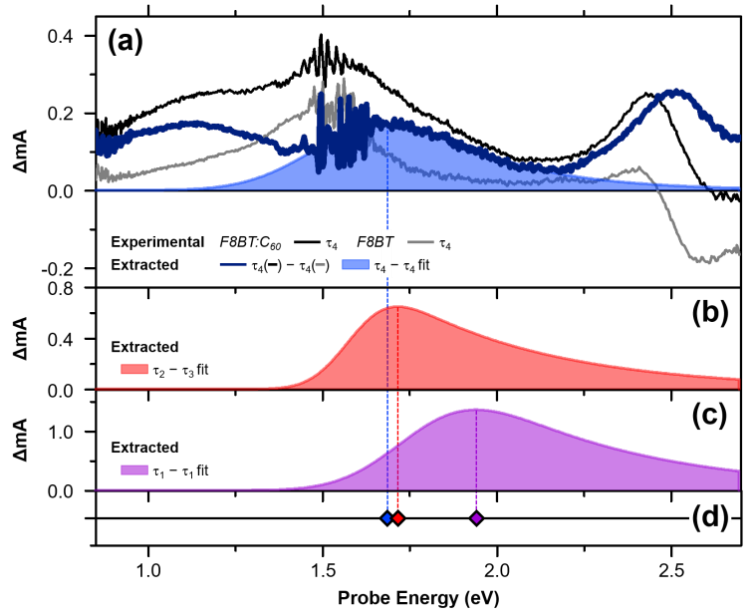

**Figure S6.** Excited-state absorption spectra corresponding to photogenerated F8BT$^{•+}$ polaron in F8BT:C$_{60}$ blend film associated with representative time constants, (a) $\tau_4$, (b) $\tau_2$, and (c) $\tau_1$. The peak position (d) shifts significantly in going from $\tau_1$ (◆; 1.94 eV) to $\tau_2$ (◆; 1.72 eV), but only marginally from $\tau_2$ (◆; 1.72 eV) to $\tau_4$ (◆; 1.68 eV).

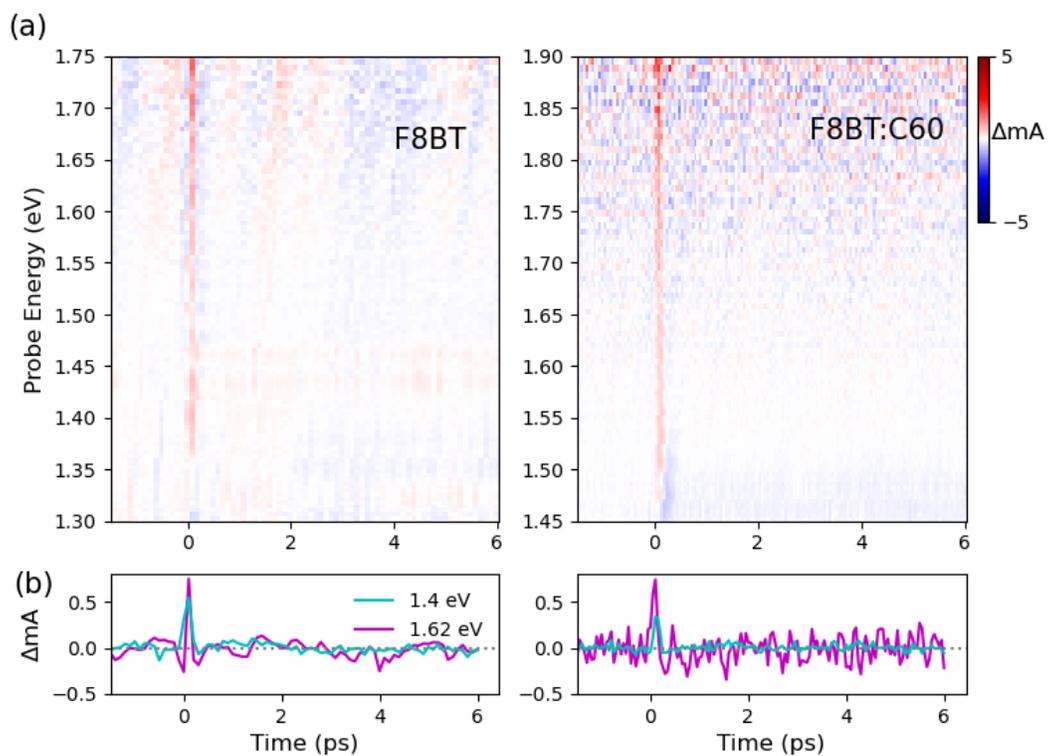

**Figure S7.** Push-probe maps as a function of push-probe delay (horizontal axis) and probe energy (vertical axis) on F8BT pristine (left panel) and F8BT:C$_{60}$ blend film (right panel). The photon energy of the push pulse is 210 meV and its fluence 450 μJ cm$^{-2}$. (b) Time traces of the maps in (a) for two selected probe energies indicated in the legend.

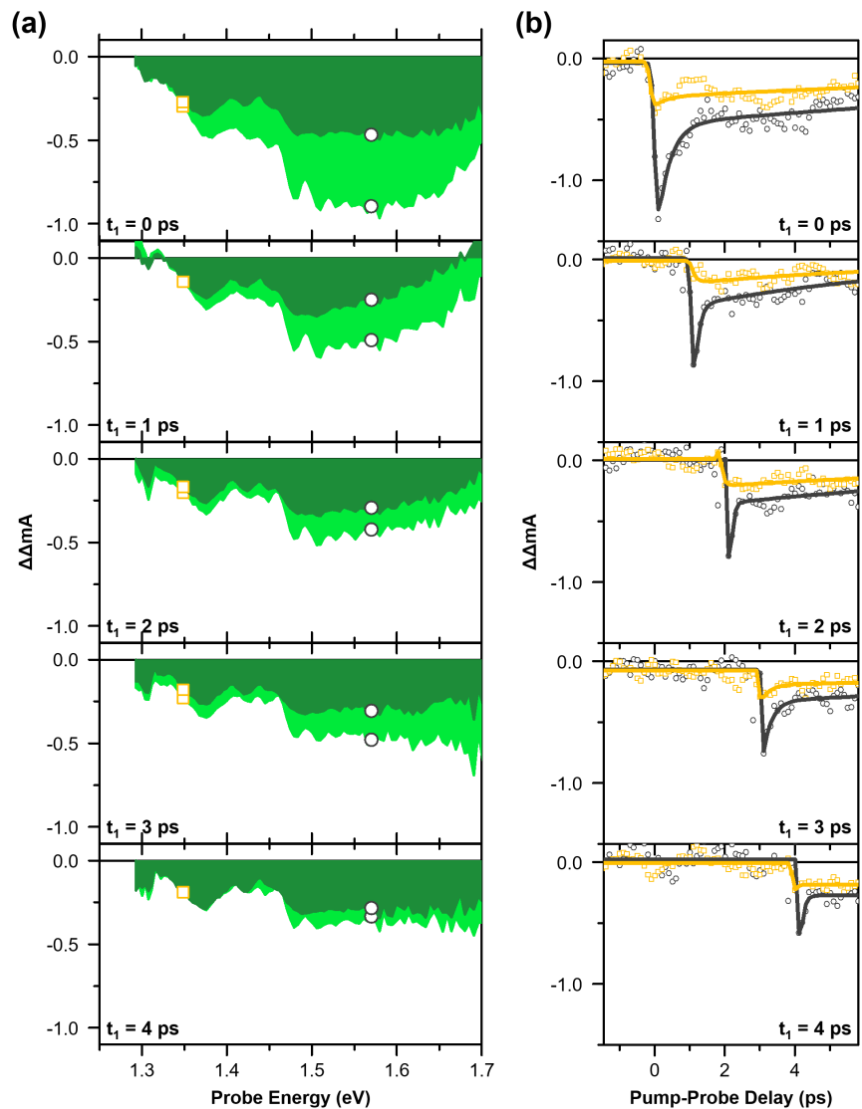

**Figure S8.** ΔΔ*A* (a) spectra and (b) time traces of F8BT pristine film as a function of pump-push time delay $t_1$. ΔΔ*A* spectra in light green shade (■) and dark green shade (■) are obtained by integrating push-probe time delay of 0.1−0.8 and 0.8−5.8 ps, respectively. The time traces are selected at representative probe energies indicated by the symbols in the spectra (○, 1.57 eV; □, 1.35 eV).

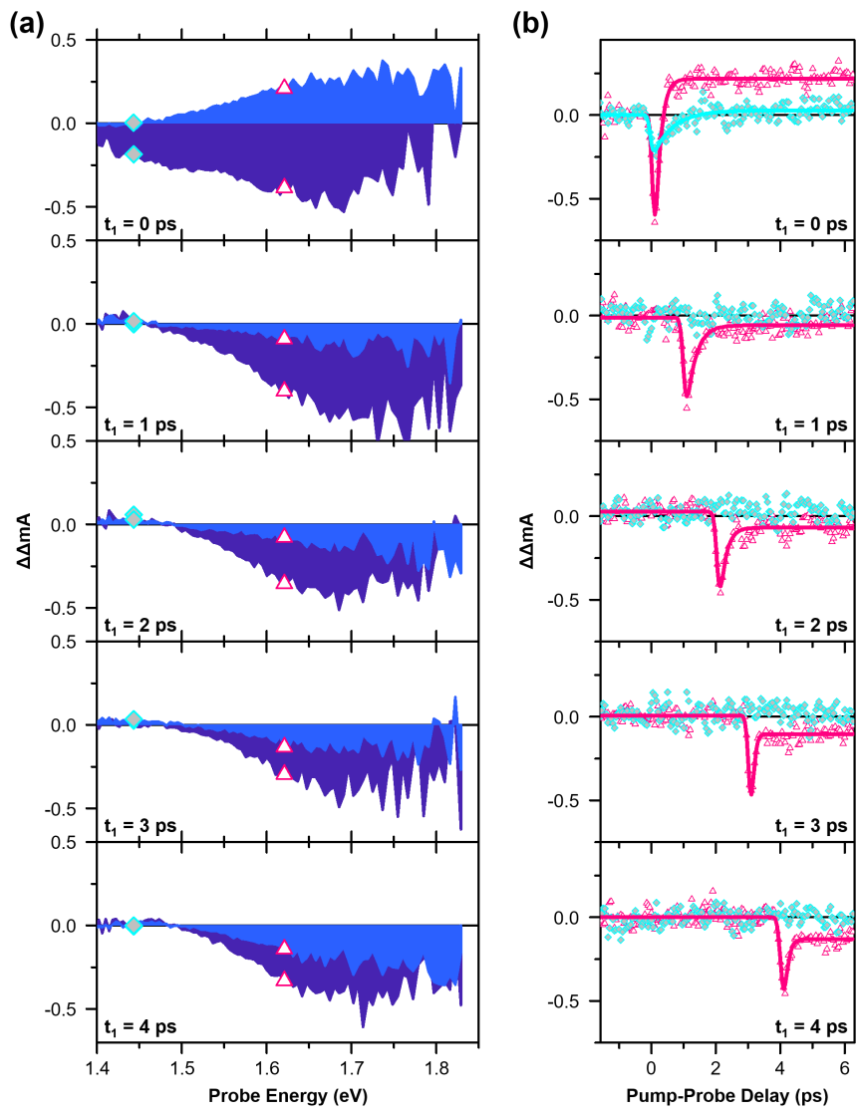

**Figure S9.** ΔΔ$A$ (a) spectra and (b) time traces of F8BT:C$_{60}$ blend film as a function of pump-push time delay t$_1$. ΔΔ$A$ spectra in blue shade (■) and purple shade (■) are obtained by integrating push-probe time delay of 0.1−0.3 and 0.9−5.8 ps, respectively. The time traces are selected at representative probe energies indicated by the symbols in the spectra (△, 1.62 eV; ◇, 1.44 eV).

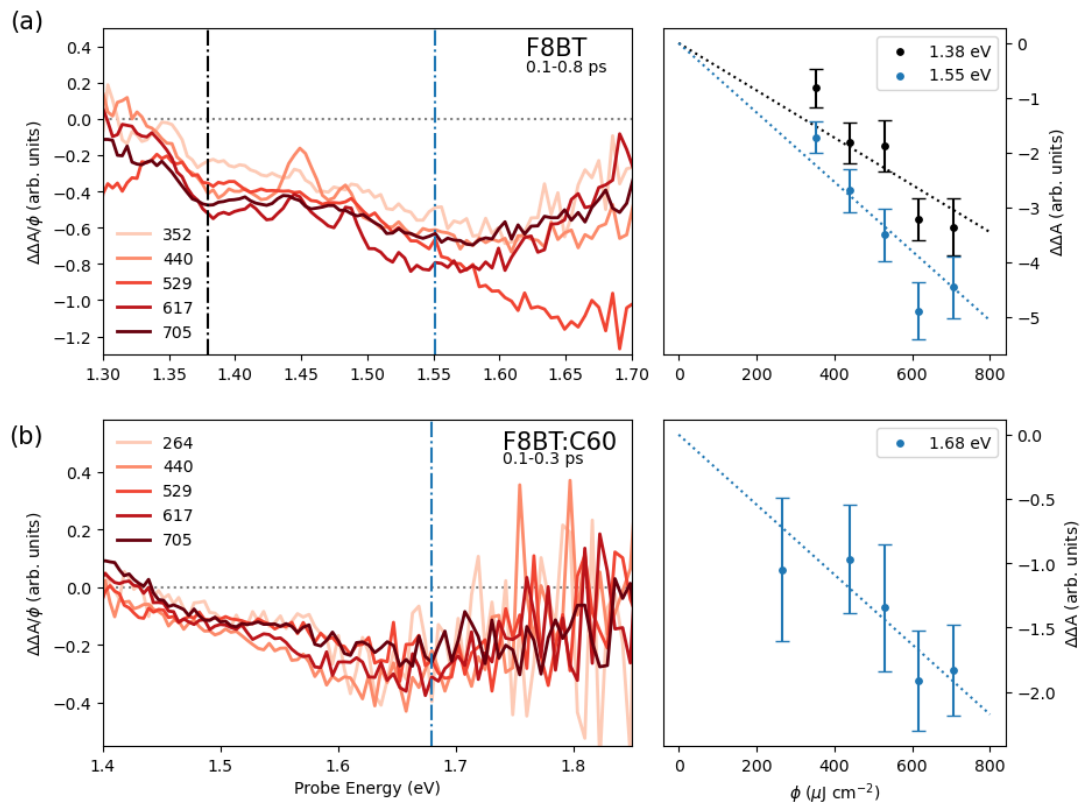

**Figure S10.** (a) Push-induced difference spectra (left panel) measured on the pristine F8BT film at different push fluences ϕ (indicated in the legend in units of µJ cm$^{-2}$). Each spectrum is integrated over the push-probe delay interval 0.1-0.8 ps and is divided by the corresponding fluence. In the right panel, the push-induced difference signal is plotted as a function of the push fluence for two selected probe energies (1.38 and 1.55 eV), marked in the left panel by the colored vertical lines. The dotted lines are linear fits to the two datasets. The error bars represent the standard deviations calculated for the selected probe energy over 5 repetitions. (b) As in (a), but for the F8BT:C$_{60}$ blend film. The spectra are integrated over the push-probe interval 0.1-0.3 ps.

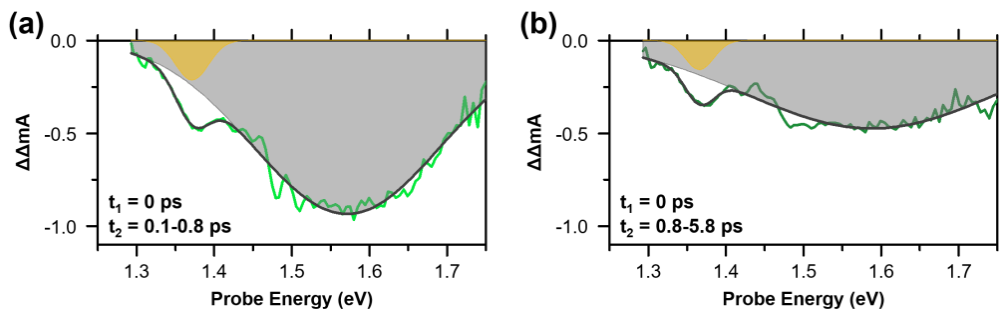

**Figure S11.** ΔΔ*A* spectra of F8BT pristine film at pump-push delay $t_1 = 0$ ps and push-probe delay $t_2 =$ (a) 0.1−0.8 ps (light green line) and (b) 0.8−5.8 ps (dark green line). Both spectra are fitted using two Gaussian peak functions (peak 1 and 2 in yellow and gray shades; sum of peak 1 and 2 in gray line). Center positions of both peaks (peak 1, 1.37 eV; peak 2, 1.58 eV) are invariant over push-probe delay $t_2$.

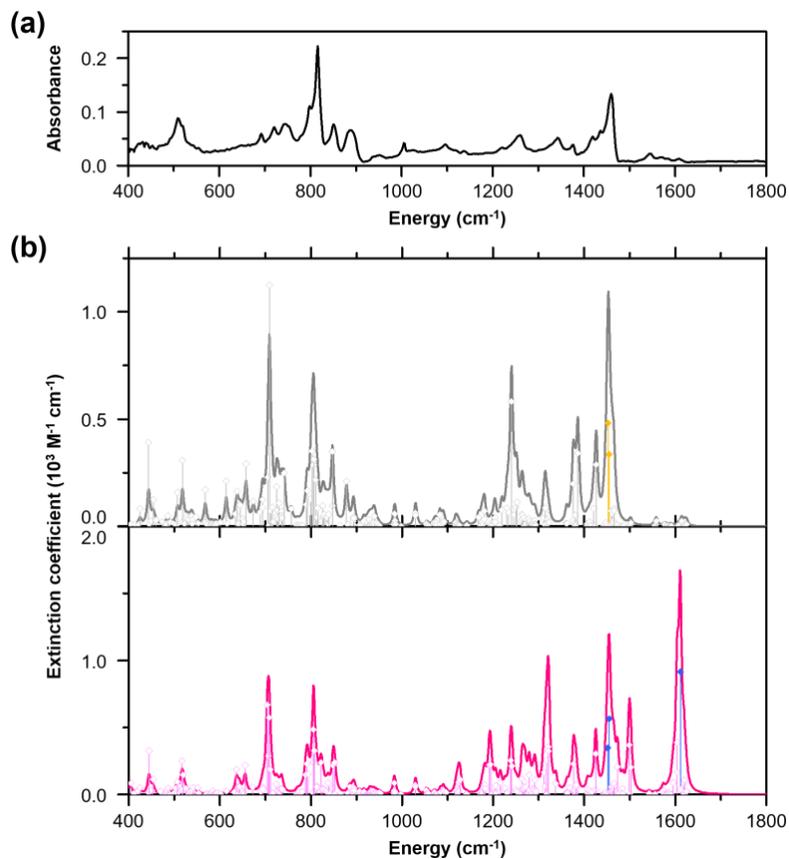

**Figure S12.** Steady-state infrared absorption spectrum of (a) neutral F8BT measured using ATR FT-IR spectroscopy. (b) Neutral (upper panel) and radical cation (lower panel) F8BT trimers (optimized structures shown in Figure S13 and S14) were used to calculate infrared extinction coefficients. The infrared spectra (solid line) are simulated by adding broadening of 8 cm$^{-1}$ FWHM to all transitions ($\diamond$ with a vertical line).

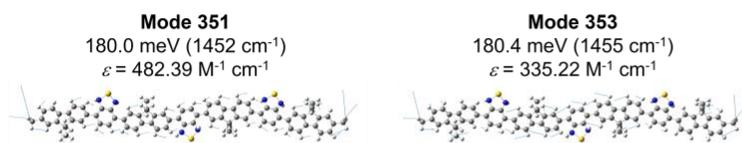

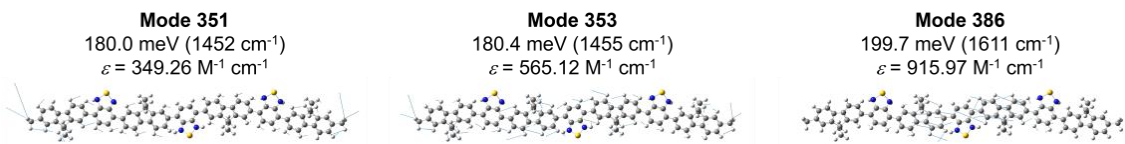

**Figure S13.** Vibrational modes that show strong push response (modes highlighted in Figure S12b). Thin blue arrows are displacement vectors of the constituent atoms.

**Singlet-singlet transitions**

**Transition 1:** 3.42 eV (363 nm), *f* = 0.0022

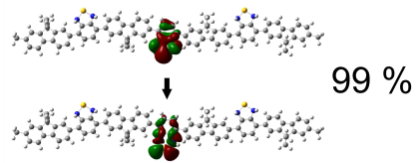

99 %

**Transition 2:** 3.94 eV (314 nm), *f* = 4.9617

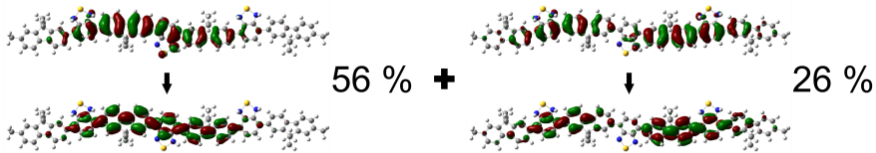

56 % + 26 %

**Doublet-doublet (radical cation) transitions**

**Transition 1:** 1.29 eV (958 nm), *f* = 0.0864

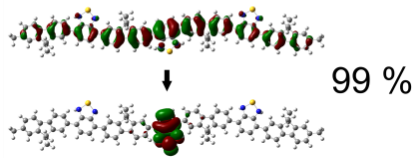

99 %

**Transition 2:** 1.44 eV (863 nm), *f* = 0.0001

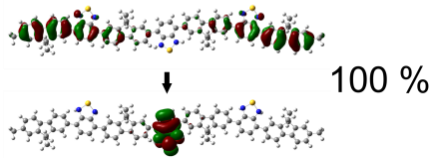

100 %

**Figure S14.** Representative natural transition orbitals (NTO) of the first two electronic transitions of neutral and radical cation F8BT trimers. Energy and oscillator strength of each transition are indicated above the NTO pair(s) that constitute the transition.

# 4. References for supporting information